# Pressure decoupled magnetic and structural transitions of the parent compound of iron based "122" superconductors BaFe$_2$As$_2$

Junjie Wu[a,b], Jungfu Lin[b *], Xiancheng Wang[a], Qingqing Liu[a], Jinglong Zhu[a], Yuming Xiao[c], Paul Chow[c] & Changqing Jin[a *]

[a]Institute of Physics, Chinese Academy of Sciences, Beijing 100190, China;

[b]Department of Geological Sciences, Jackson School of Geosciences, The University of Texas at Austin, TX 78712, USA;

[c]HPCAT, Carnegie Institution of Washington, Advanced Photon Source, Argonne National Laboratory, Argonne, IL 60439, USA

* The corresponding authors: afu@jsg.utexas.edu; Jin@iphy.ac.cn

**The recent discovery of iron ferropnictide superconductors has received intensive concerns on magnetic involved superconductors. Prominent features of ferropnictide superconductors are becoming apparent: the parent compounds exhibit antiferromagnetic (AFM) ordered spin density wave (SDW) state; the magnetic phase transition is always accompanied to a crystal structural transition; superconductivity can be induced by suppressing the SDW phase via either chemical doping or applied external pressure to the parent state. These features generated considerable interests on the interplay between magnetism and structure in chemical doped samples, showing crystal structure transitions always precedes to or coincide with magnetic transition. Pressure tuned transition on the other hand would be more straightforward to superconducting mechanism studies since there are no disorder effects caused by chemical doping; however, remarkably little is known about the interplay in the parent compounds under controlled pressure due to the**

**experimental challenge of in situ measuring both of magnetic & crystal structure evolution at high pressure & low temperatures. Here we show from combined synchrotron Mössbauer and x-ray diffraction at high pressures that the magnetic ordering surprisingly precedes the structural transition at high pressures in the parent compound $BaFe_2As_2$, in sharp contrast to the chemical doping case. The results can be well understood in terms of the spin fluctuations in the emerging nematic phase before the long range magnetic order that sheds new light on understanding how parent compound evolves from a SDW state to a superconducting phase, a key scientific inquiry of iron based superconductors.**

# Introduction

High pressure is very effective to modify crystal or electronic structures that often lead to new materials states of interdisciplinary interests from geosciences across chemistry to physics [1~5]. Pressure is especially unique in generating or tuning superconducting states with the advantages without introducing disorder effects in comparing to chemical doping [3], hence more favorable to intrinsic properties studies. The recent discovery of iron pnictide (or ferropnictides) superconductors [6 ~ 9] has shed new light on unconventional superconductors involved with magnetic elements that are previously considered to be incompatible to superconductivity. The spins of iron in ferropnictides are periodically modulated and localized in space forming the spin density wave (SDW) parent state, whereas such antiferromagetically-ordered magnetism (AFM) is also closely related to the superconductivity in cuprates and other unconventional superconductors. Magnetism has been suggested to play a crucial role in the occurrence of high $T_C$ in ferropnictides, in which the paramagnetic (PM) to AFM transition is accompanied by a tetragonal to orthorhombic structural transition in systems like LaFeAsO (10) or $BaFe_2As_2$ (11).

The "122" ternary $AFe_2As_2$ (A = alkaline metals Ca, Sr and Ba) compounds, characterized by pairs of tetrahedral FeAs layers, have clearly demonstrated pressure-induced high $T_C$ of superconductivity with well-defined magnetic and structural characteristics (12~15). Upon cooling, the undoped parent compounds of the "122" system exhibit a tetragonal to orthorhombic structural transition strongly coupled by a paramagnetic to antiferromagnetic transition that takes places at the temperature in coincide with the structural phase transition (14). Chemical doping has been found to strongly suppress this transition. Recent study on $Ba(Fe_{1-x}Co_x)_2As_2$ shows that the structural transition precedes the magnetic transition in decreasing temperature for a fixed amount of chemical doping fraction (16). Moreover, experimental results on underdoped $Ba_{0.84}K_{0.16}Fe_2As_2$ single crystals, with excellent chemical homogeneity, have further shown that the magnetic and structural transitions are indeed separated (17) with the structural transition occurring at higher temperature. It is noted that the antiferromagnetic transition is always preceded by or coincident with a tetragonal to orthorhombic structural distortion. This observation indicates that the structural distortion is related to an electronic phase transition due to C4 symmetry break as a result of long range ordering (18, 19 20).

High pressure has been shown to effectively suppress AFM ordering and to induce a high $T_C$ in the system (21,22). Similar to the chemical doping, applied pressure has been found to cause the same effects on tuning the SDW state to superconductivity (23~28). Since AFM ordering is strongly coupled with structural distortion, further investigation under controlled *pressure temperature (P-T)* conditions using *in situ* probes sensitive to these transitions, is needed to unravel the intrinsic interplay between structure and magnetism in the parent compound of iron based superconductors. Here we report the studies on the magnetic and structural transitions of $BaFe_2As_2$ investigated using synchrotron Mössbauer spectroscopy (SMS) and x-ray diffraction (XRD) in a cryogenically-cooled high-pressure diamond anvil cell (DAC) with highly hydrostatic pressure medium of neon. Study of the "122" system at high pressures was chosen as high-quality single crystal and powder samples can be well synthesized and prepared. Our SMS spectra with very high energy resolution were used to derive hyperfine parameters that are extremely sensitive to the local magnetic moments ordering of the sample, while micro focused XRD patterns provide information on lattice parameters and their degrees of distortion. The combined results show that mesoscopic magnetic ordering that can be detected by subtle SMS precedes the structural distortion over temperature cooling at high pressures, different from that of chemical doping where the crystal structure transitions is always proceeds the magnetic phase transitions. These results indicated the emerging nematic phase before long range spin order.

## Results & Discussion

The temperature-dependent SMS spectra for $BaFe_2As_2$ at various pressures are well represented by the modeled SMS spectra (Fig. 1). The higher temperature tetragonal phase exhibits such that its SMS spectrum is shown as a smooth, exponentially-decayed line (29). Over temperature decrease at high pressures, $BaFe_2As_2$ undergoes a magnetic transition with the occurrence of the hyperfine magnetic moments shown as the first minimum quantum beat in the SMS spectra (Fig. 1). Our derived hyperfine magnetic field is consistent with previous results under ambient pressure and low temperatures (30); at 100 K and ambient pressure, our calculated hyperfine field is 5.1 ($\pm$1.0) T as compared to 5.23 T observed at 77 K in the literature (30). The derived hyperfine magnetic fields as a function of temperature show the onset of the magnetic transition with a sharp increase in the magnetic moment (Fig. 2). As pressure increases, the hyperfine magnetic field decreases and the magnetic transition region broadens, implying the suppressed magnetism by applied pressure. The magnetic transition pressure in the powder sample is slighter lower than that in the single-crystal sample at a given temperature, indicating that enhanced anisotropic strain decreases the magnetic transition pressure.

Rietveld full-profile refinements of the micro focused XRD spectra shows the tetragonal (*I4/mmm*) to orthorhombic (*Fmmm*) structural transition with decreasing temperature at 12.5 GPa (Fig. 3A). The (220) line can be well modeled for a single peak at 80 K in the tetragonal phase, while it is evident that the (220) line splits into (400) and (040) peaks at 15 K. The variation of the lattice parameters with decreasing temperature is depicted in Fig. 3B. As the splitting of the (220) peak in the tetragonal structure to the (400) and (040) peaks in the orthorhombic structure at 30 K, the *a* axis splits into *a* and *b* axes, showing the onset of the structural transition in the *P-T* phase diagram (Fig. 3, 4). We also show a portion of the diffraction patterns of $BaFe_2As_2$ at selected pressures at 15 K with 2θ between 14.3° and 20° in Fig. 3C. The (400) and (040) peaks merged into a single peak at around 13.7 GPa, indicating that applied pressure drives the structural transition from the tetragonal to the orthorhombic phase at 15 K.

Using both high-resolution SMS and XRD results, we have constructed the phase diagram of $BaFe_2As_2$ as function of high pressures and low temperatures (Fig. 4). The extremely energy sensitivity plus high spatial resolution based on synchrotron Mossbauer Effects at high pressure provides the possibility to detect mesoscopic ordered local magnetic moment, i.e., the nematic phase that will be further addressed in the discussion part. Thus obtained magnetic order temperature is assigned to $T^*$ as shown in Fig. 4, whereas the structural transition temperature ($T_S$) was derived from the refined XRD patterns. Since the magnetic transition occurs over a certain temperature region at a given pressure, the onset of the $T^*$ is derived as the mean of the two experimental temperatures between the last appearance of the PM state and the first appearance of the AFM state (Figs. 2, 4). For comparison, the superconducting temperature ($T_C$) was taken from previous electrical resistivity measurements using hydrostatic medium similar to our experiments (28). Below 3 GPa, the magnetic transition occurs over a narrower temperature region and the $T^*$ curve coincides with the $T_S$ by XRD and resistivity measurements as shown in Figs. 2 & 4. At higher pressures, the magnetic transition broadens and the onset of the $T^*$ curve increasingly precedes the $T_S$ curve in temperature decrease even up to about 15 GPa, where the magnetic ordering has been completely suppressed.

The separation and deviation of the structural and magnetic transitions, has been reported in "1111" type LaFeAsO and NdFeAsO, with the structural transition preceding the magnetic transition (31, 32). However, in the peculiar 122 type parent compounds, the separation of the structural and magnetic transitions is only observed in chemically doped compounds with induced disorder, in which the structural transition is observed to occur before the magnetic transition in temperature decrease. Here we use the nematic magnetic order model to explain the separation of the transitions at high pressures (33). The nematic transition temperature, which describes the orientation order between two different AFM sublattices, occurs above the onset of the static magnetic order, and has been observed in chemical doped $BaFe_2As_2$ compound (34,35). For the iron arsenide compounds, the nematic phase corresponds to the orientation order between two AFM sublattices that are only weakly coupled in the PM state because of the frustration induced by the next-nearest-neighbor magnetic interactions (36). With the application of high pressure, an increased alignment of the magnetic moments is induced before the structural transition takes place, indicating that localized AFM alignment of the spins can precede the structural transition. This temperature corresponds to the $T^*$. The magnetic transition encourages the structural phase transition from the tetragonal to the orthorhombic phase under high pressure via the C4 symmetry broken. The $BaFe_2As_2$ compound has been proposed to be a dynamic SDW phase with a stripe AFM state (37). According to the model, the interlayer magnetism along the $c$ axis is very small, and the anti-phase boundaries are formed along the $a$ and $b$ axes of the orthorhombic phase at low temperatures. Considering that the magnetic transition concurs with the structural transition in the $BaFe_2As_2$ system at ambient pressure, the anti-phase boundaries are pinned by the interlayer coherency along the $c$ axis below the transition temperature. Above the structural transition temperature, the twinning domain will dominate the PM state. It should be noted that, on the timescale of our SMS experiments in the order of $\sim 10^{-8}$ s, the fluctuations of the twinning domains (37) (the nematic phase) are too fast to be observed completely. In fact what we have observed by the high-resolution Mössbauer should be the local spin that occurs before the macroscopic structural phase transition. Although the orientation of the twinning domains is conserved in the FeAs layers across the structural transition at low temperatures, the bulk magnetic symmetry breaks apart from the C4 symmetry. As a result, the AFM state exhibits detectable magnetic moments shown in Mössbauer spectra in three dimensions, and it occurs in decreasing temperature before the structural transition under high pressures.

Our results here indicate that $Fe^{+2}$ exhibits mesoscopic spin moments right before the structural transition in temperature decrease. That is, the local moments involving the fluctuation and interaction of the electron spins thus play a key role in the origin of superconductivity in pnictides. These results enable one to understand how the magnetic order in a mesoscopic scale proceeds to the macroscopic crystal structure phase transition in high $T_C$ ferropnictide. These high-pressure results reveal a pure tuning on the physics of the system that is distinct from the chemical doping, providing valuable, direct insight into exploring the interplay between magnetism and structure transitions.

## Methods

**Single Crystal Growth**

Single crystalline $BaFe_2As_2$ samples with 20 atom% $^{57}Fe$ enrichment were grown by the FeAs self-flux method (15).

**Magnetic moment detection at high pressures**

Powder samples were gently ground from the $^{57}Fe$-enriched single crystals, and were later used for SMS and x-ray diffraction experiments. SMS experiments were conducted in a membrane DAC at the beamline 16-IDD of the Advanced Photon Source (APS). A synchrotron x-ray beam of 30 (V) × 50 (H) μm, $\Delta E/E \sim 1*10^{-7}$ in resolution and 14.4125 keV corresponding to the nuclear transition energy of the $^{57}Fe$ nuclei was used to measure the SMS spectra of $^{57}Fe$-doped sample. Both powder and single crystal samples with 100 μm in diameter and 15-20 μm in thickness (corresponding to an effective thickness of five) were used for the experiments in order to understand the effect of anisotropic stress on the magnetic transition. A rhenium gasket was pre-indented to a thickness of 30 μm by a pair of 400 μm culets, and a hole of 150 μm was subsequently drilled in it and used as the sample chamber. The sample was loaded into the sample chamber using Ne as the pressure medium, together with a few ruby spheres for pressure determination. A helium-cooled cold-finger cryostat was used to cool the sample in the DAC to as low as 6 K; temperatures of the DAC were equilibrated for a couple of hours after each cooling and were measured using two thermocouples attached each side of the DAC. Pressures of the sample chamber were controlled by a membrane diaphragm and measured *in situ* using an online ruby system before and after the SMS collection. To determine the effects of hydrostaticity on the magnetism of the sample, the SMS experiments were also conducted using mineral oil (a mixture of heavy hydrocarbons) as a pressure medium. SMS spectra were collected by an avalanche photodiode detector at the nuclear forward scattering geometry with a collection time of 1-2 hours. The SMS spectra were fitted using the CONUSS(COherent NUclear resonant Scattering by Single crystals) program (38) to derive the hyperfine magnetic fields.

**Structure characterizations at high pressures**

Angle-dispersive powder XRD measurements were carried out at the 16-IDD beam line of the APS. A monochromatic X-ray beam with a wavelength of 0.37379 Å and a focused beam size of 5 (V) × 10 (H) μm was used. The $^{57}$Fe-enriched powder samples were loaded into the same DAC in the same cryostat using a Ne pressure medium and ruby pressure calibrate. Analyses of the XRD patterns, before and after the grinding, showed consistent sample lattice parameters, ensuring that grinding does not introduce additional strain on the sample. The powder XRD measurements were performed at pressure up to 15 GPa from room temperature to 15 K with a typical temperature interval of 10 K. Rietveld full-profile refinements of the XRD pattern were performed using the GSAS package. The use of the same sample batch with similar experimental conditions (Ne pressure medium, ruby scale, DAC, and cryostat) and small x-ray beam sizes permits reliable comparison on the magnetic and structural transitions.

**ACKNOWLEDGEMENTS**

We thank C. Kenney-Benson, J. Liu, and C. Lu for their assistance; Z. Mao for helping with data analyses; and A. Wheat for editing the manuscript. We are also grateful to Wenge Yang, Yusheng Zhao, and Guoyin Shen for discussions. Portions of this work were performed at the High Pressure Collaborative Access Team (HPCAT Sector 16), Advanced Photon Source, Argonne National Laboratory. HPCAT operations are supported by the National Nuclear Security Administration, Department of Energy under Award DE-NA0001974 and by the Basic Energy Sciences, Department of Energy (DOE-BES) under Award DE-FG02-99ER45775, with partial instrumentation funding by the National Science Foundation. Advanced Photon Source, Argonne National Laboratory is supported by DOE-BES under Contract DE-AC02-06CH11357. Work at The University of Texas at Austin is supported by Energy Frontier Research in Extreme Environments (EFree) and the Carnegie/DOE Alliance Center. Work at the Chinese Academy of Sciences is supported by National Science Foundation and Ministry of Science and Technology of China through research projects.

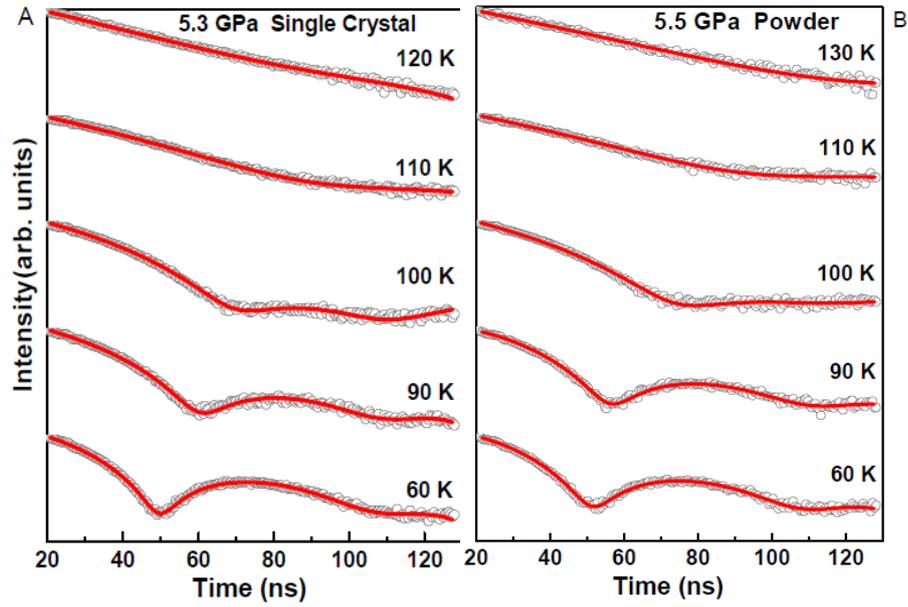

**FIG. 1** Representative temperature-dependent SMS spectra of BaFe$_2$As$_2$ in Ne pressure medium. (A) Single crystal at 5.3 GPa. (B) Powder sample at 5.5 GPa. Open circles: experimental data; red solid lines: modeled spectra using CONUSS program. Quantum beats in the spectra represent the occurrence of the magnetic fields and hence the AFM state, whereas the absence of beats (flat spectral feature) indicates the PM state.

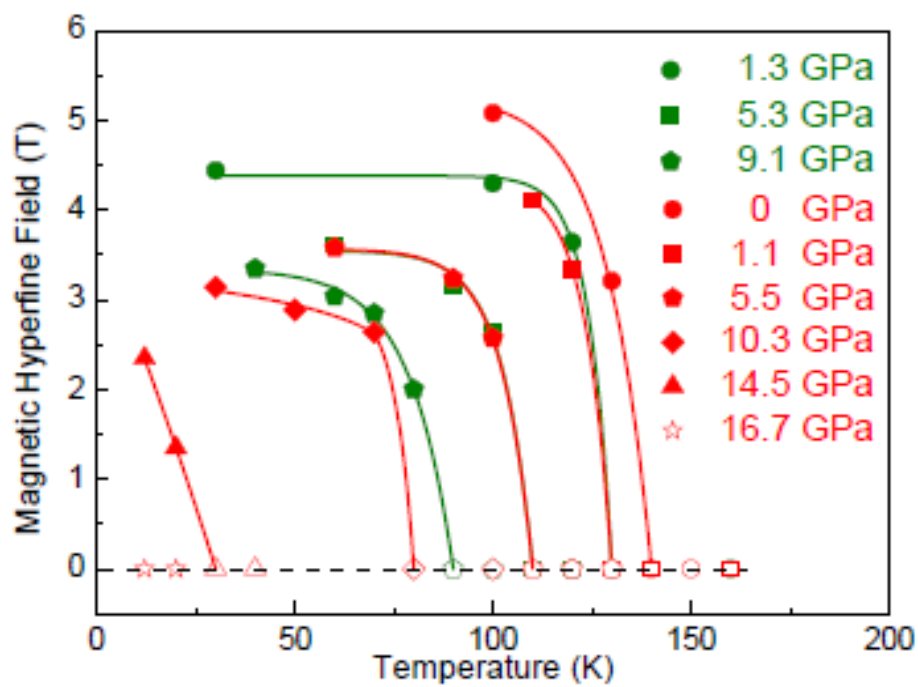

**FIG. 2** Derived hyperfine magnetic fields of BaFe$_2$As$_2$ as a function of temperature at high pressures. Red symbol: powder in Ne medium; olive symbol: single crystal in Ne medium; filled symbol: AFM state; open symbol: PM state. Solid lines: simple fitting to the data.

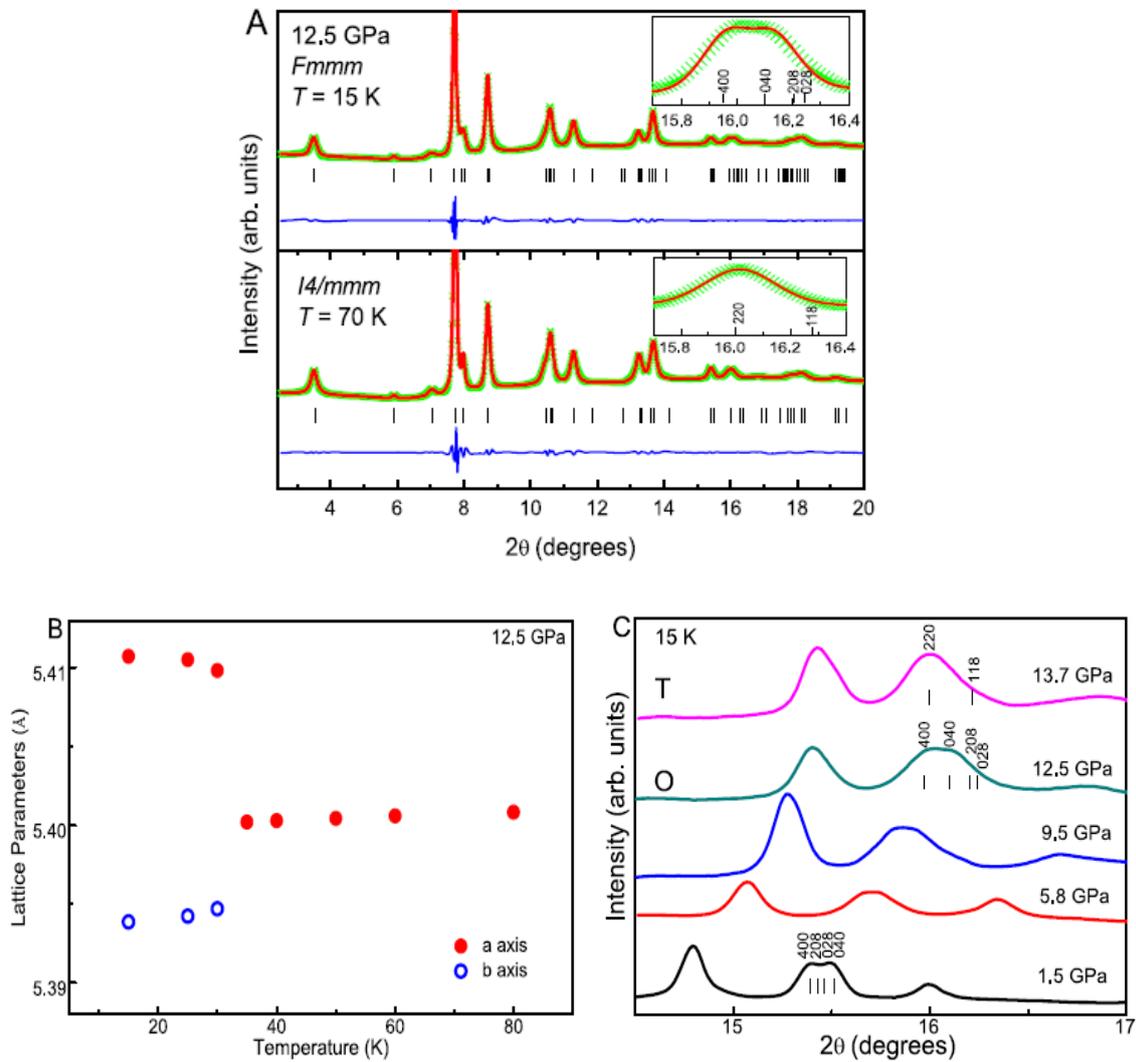

**FIG. 3** Representative synchrotron XRD patterns of BaFe$_2$As$_2$ at high pressures. (A) Rietveld refinement of the measured pattern at 15 K (*Fmmm*) and 80 K (*I4/mmm*) under 12.5 GPa. Inset shows the splitting of (220) reflection to (400)/(040). (B) Variation of the lattice parameters with decreasing temperature at 12.5 GPa. (C) Pressure-dependent powder synchrotron x-ray diffraction patterns of BaFe$_2$As$_2$ at 15 K. Ort, Orthorhombic structure; Tet, tetragonal structure;

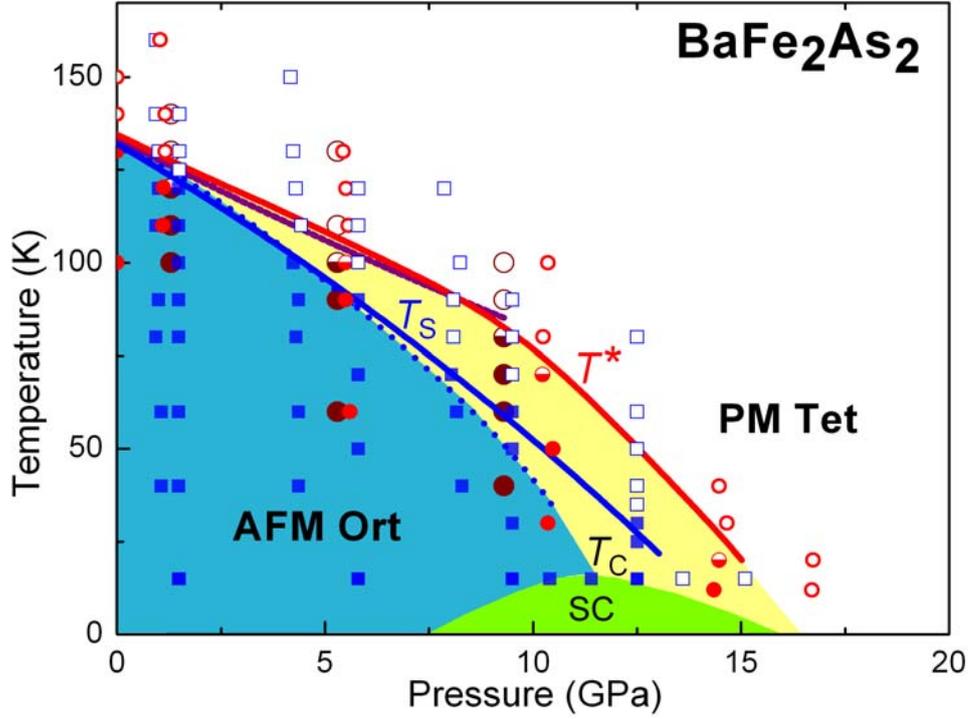

**FIG. 4** Schematic P-T phase diagram of BaFe$_2$As$_2$. Red symbols: powder SMS results; brown symbols: single crystal SMS results; blue symbols: XRD data. Red solid line: $T^*$ (nematic phase temperature) from SMS (Ne, powder); brown solid line: $T^*$ from SMS (Ne, single crystal); blue solid line: $T_S$ from XRD (Ne, powder); $T_C$, blue dotted line & $T_S$ from Ref 28. SC: superconducting phase; Tet: tetragonal structure; Ort: orthorhombic structure; AFM: antiferromagnetic; PM: paramagnetic. Pressure uncertainties are smaller than the symbols, whereas temperature uncertainties are less than 5 K.